# Experimental demonstration of Tessellation Structured Illumination Microscopy

Doron Shterman, † and Guy Bartal †*

† The Erna and Andrew Viterbi Faculty of Electrical and computer Engineering, Technion – Israel Institute of Technology, Haifa 32000, Israel

**ABSTRACT:** Structured Illumination Microscopy (SIM) overcomes the optical diffraction limit by folding high-frequency components into the baseband of the optical system, where they can be extracted and then repositioned to their original location in the Fourier domain. Although SIM is considered superior to other super-resolution (SR) methods in terms of compatibility with live cell imaging and optical setup simplicity, its reliance on image reconstruction restricts its temporal resolution and may introduce distortions in the super-resolved image. These inherent drawbacks are exacerbated in extended-SIM implementations, where spatial resolution surpasses the diffraction limit by more than 2-fold. Here, we present and demonstrate the *Tessellation Structured Illumination Microscopy* (TSIM) framework, which introduces a revived image reconstruction paradigm. With TSIM both the temporal resolution limit and the reconstruction artifacts that impact extended-SIM, are alleviated, without compromising the achievable spatial resolution. Keywords: Super-resolution, fluorescence microscopy, structured illumination.

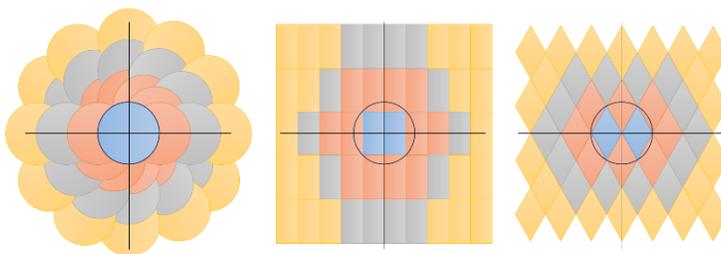

The ultimate temporal resolution limit in super-resolution (SR) fluorescence microscopy is dictated by the fluorophore excitation-emission cycle. In practice, this limit has never been reached owing to system-level constraints like physical apparatus, operation concept, detector efficiency, and image acquisition time. In single-molecule localization methods, such as STORM[1] and PALM[2], the temporal resolution is set by signal-to-noise (SNR) constraints which are met by capturing multiple frames to locate individual molecules and reveal their structures. In patterned-illumination methods, like parallelized STED[3] and SIM[4], the temporal resolution is limited by either the scanning duration (OL-STED[5]) or by the need to acquire multiple frames to resolve frequency aliasing and reconstruct a super-resolved image. A variety of strategies have been employed to enhance the temporal resolution in SR microscopy[6,7], though a trade-off between spatial and temporal resolution persists.

Currently, SIM is considered among the fastest SR microscopy techniques[8] due to its wide field of view and superior photon efficiency[9], where most of the photons collected are used for super-resolved image formation. Reducing the overall number of frames captured in SIM has been suggested[10,11] to further improve its temporal resolution and reduce phototoxicity effects, allowing live-cell SR imaging. However, these frame reduction methods rely on complicated reconstruction algorithms with high sensitivity to the reconstruction parameters, resulting in significant artifacts in the super-resolved image. Reducing the total number of frames is essential for data-intensive SIM techniques, such as MM-SIM[12] and nonlinear SIM[13], which, while offering up to 5-fold increase in spatial resolution, require a staggering 72 frames per single super-resolved image.

Here, we introduce the Tessellation Structured Illumination Microscopy (TSIM) framework, allowing robust, continuous expansion of the Fourier domain in extended-SIM, improving spatial resolution by over 2-fold beyond the diffraction limit. This is achieved by replacing the conventional high-frequency folding paradigm, where circular regions in the Fourier domain are concatenated symmetrically[14], with polygonal regions and precise selection of frequency folding centers. Combined with a cascading reconstruction algorithm[15] that leverage prior knowledge regarding shifted frequency components, TSIM offer a robust and efficient super-resolution reconstruction. With TSIM, the temporal resolution of extended-SIM[16-18] improves by up to 3-fold.

The optical resolution limit, represented by the cutoff frequency $k_0$ in Fourier domain, Figure 1a, is the maximum spectral frequency resolved using an optical system:

$$k_0 = \frac{2NA}{\lambda} = \frac{2n \cdot \sin(\alpha)}{\lambda} \quad (1)$$

where: $\lambda$ is the wavelength of light, $NA$ is the numerical aperture of the imaging system, $n$ is the index of refraction, and $\alpha$ is the half angle of the maximum cone of light that can enter or exit the lens.

The optical diffraction limit can be described as a low-pass filter in the Fourier domain, allowing only frequencies below the cutoff to pass. Information encompassed in frequencies beyond this limit cannot be resolved, causing images with such high frequencies to appear blurred or dim. Using patterned illumination high-frequency spectral content can be recovered to create a super-resolved image.

Recalling the SIM framework, an illumination profile generated with two coherent beams takes the form of:

$$I_{ill}(r) = I_0 \left[ 1 + \frac{\cos}{2}(2\pi p \cdot r + \varphi) \right] \quad (2)$$

where $r \equiv (x, y)$ is the two-dimensional spatial position vector, $I_0$ is the peak illumination intensity taken as unity for simplicity from this point forward, $p$ is the illumination frequency vector in the Fourier domain, and $\varphi$ is the phase of the illumination profile.

Using the Fourier theorem, the emitted light is given by:

$$\tilde{I}_{em}(k) = \tilde{s}(k) \otimes \tilde{I}_{ill}(k) \quad (3)$$

where $k \equiv (k_x, k_y)$ is the two-dimensional Fourier-domain position vector, $\tilde{s}(k)$ is the fluorophore density distribution in the Fourier domain, and $\otimes$ is the convolution operator.

The detected signal is given by:

$$\tilde{I}_{dec}(k) = \tilde{I}_{em}(k) \cdot \tilde{h}(k) = [\tilde{s}(k) \otimes \tilde{I}_{ill}(k)] \cdot \tilde{h}(k) \quad (4)$$

where $\cdot$ denotes point-wise multiplication and $\tilde{h}(k)$ is the system Optical Transfer Function (OTF).

Hence, for a given illumination profile phase $\varphi_m$, the detected signal in the Fourier domain can be written as:

$$\tilde{I}_{dec}(k) = \left[ \tilde{s}(k) + e^{i\varphi_m} \frac{\tilde{s}}{2}(k+p) + e^{-i\varphi_m} \frac{\tilde{s}}{2}(k-p) \right] \cdot \tilde{h}(k) \quad (5)$$

According to Equation (5), the detected signal $\tilde{I}_{dec}(k)$ contains two shifted and one non-shifted spectral component $\tilde{s}$ within the OTF passband region. In classical SIM reconstruction, the shifted spectral components can be de-aliased by capturing a set of three consequent frames while changing the illumination profile phase $\varphi_m$ in between. This process is then repeated with illumination profiles in multiple spatial orientations to cover the entire Fourier domain. De-aliasing, or spectral component un-mixing, dictates the number of frames captured in most SIM realizations.

Interestingly, prior knowledge of the spatial distribution of the shifted frequency components can be used for de-aliasing, negating the need for multiple frames per illumination orientation, as illustrated in Figure 1. For ease of annotation, we use $\tilde{s}_-$, $\tilde{s}_+$ and $\tilde{s}_0$ to denote the shifted and unshifted spectral components, respectively, and $\tilde{s}_{ij}$, with i = $[0, -, +]$ and j = $[L, R]$, to denote the subset spectral components residing on the left or right side of the y-axis. Because spectral components residing within the $\tilde{s}_0$ region can be directly accessed using a simple diffraction-limited image, only one additional frame capture is required to de-alias the remaining $\tilde{s}_{-L}$ and $\tilde{s}_{+R}$ components. Therefore, by using one diffraction-limited image and one extra frame for each illumination direction, a super-resolved image can be reconstructed with approximately three times fewer frames than in classical SIM reconstruction.

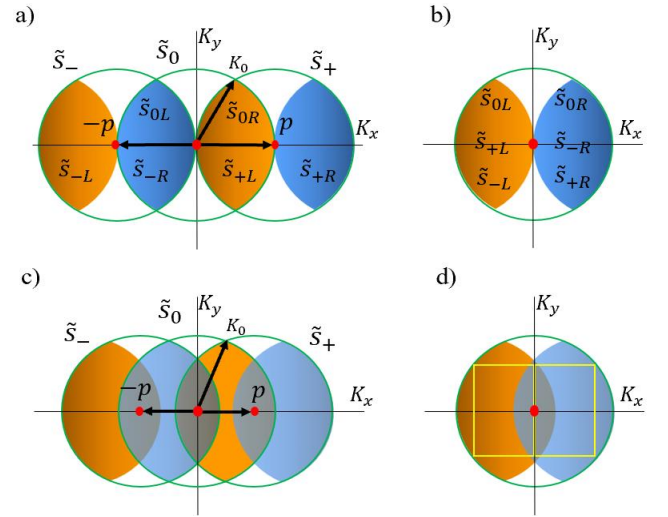

Figure 1: Geometrical illustration of aliasing in SIM. $k_0$ is the diffraction cutoff frequency and $p$ is the illumination profile frequency. (a) 1D spectral components in SIM for $p=|k_0|x$, prior to aliasing. Three distinct spectral regions, being the DC and two shifted components, are encapsulated within green circles, with cutoff frequency $k_0$ corresponding to the 2D OTF of a circular aperture. Blue and orange colors, contained within an ellipse-like shape, represent the overlapping areas between the DC and shifted spectral components. (b) Spectral components distribution in a post aliasing configuration. (c) 1D spectral components in SIM for $p<|k_0|x$ before aliasing and after in (d) showing larger ellipse-like overlapping regions. The yellow box in (d) highlights polygonal regions that can be utilized for the practical and robust tessellation reconstruction process.

This frame-reduction technique eliminates the need to control the phase of the illumination profile, simplifying the optical setup and improving reconstruction robustness[19]. The overlapping areas between two OTF shapes, shown as ellipse-like regions in Figure 1, dictate the recoverable spectral regions. However, two degrees of freedom remain: first, the illumination frequency $p$ can be adjusted, and second, a subset of the overlapping regions can be cropped out either optically or digitally. TSIM leverages these degrees of freedom to expand the reconstruction mechanism in a cascading manner, enabling a practical and efficient reconstruction. We consider tessellation, or tiling, as the continuous expansion of the Fourier domain using one or more geometric shapes, without gaps or overlaps. While geometric tessellation typically obeys pre-defined mathematical rule set[20] to comply with symmetry or topology constrains, in TSIM the tessellation must comply with constraints arising from reconstruction constrains, such as enabling frequency unmixing. To accomplish this, the tessellation must cover every point in the Fourier domain as many times as spectral components have been folded together. Another restriction is the OTF shape, which dictates both the observable and folded Fourier regions, therefore the selected tiles must reside within this shape. Lastly, since most image-sampling devices and digital image representations use rectangular pixel arrays, polygonal tile shapes are preferred. This can simplify the reconstruction process and reduce artifacts compared to the use of curvilinear shapes.

Given these constraints, multiple tessellation schemes can be evaluated in terms of spatial resolution, temporal resolution, and data collection efficiency. In this work, we implement a rectangle shape as the baseline tile for tessellation and demonstrate TSIM in an MM-SIM equivalent configuration, both numerically and experimentally. Our demonstration relies on accurate control of the Illumination profile to select the frequency centers around which high-frequency content is being folded.

The 2D OTF of an imaging system with a circular aperture has a circular shape in the Fourier domain, Figure 2a, with a cutoff frequency of $k_0$. Traditionally, Fourier domain expansion in nonlinear SIM or extended linear SIM is achieved by combining the circular shaped OTF region, being isotopically distributed in Fourier domain, Figure 2b. Instead, using polygonal tiles like rectangles or rhombuses, Figure 2c-d, the total number of frames is reduced, since the overlap between folded frequency components is minimized. These polygonal tiles can be cropped out optically or digitally from within the 2D circular OTF. To ensure overlap with the known frequency content, the selected tile shape must not only fit within the OTF shape but also be contained within the ellipse-like overlapping area, as shown in Figure 1.

Other, more complex combinations of different polygonal and non-polygonal[21] shapes can also be considered.

Comparing the number of frames required for extended linear SIM, we can observe the data collection efficiency provided by TSIM. Classical MM-SIM reconstruction for an optical system with a circular OTF, Figure 2a, requires 18 discrete frequency folding centers, each sampled three times, totaling 54 frames, Figure 2b. For the same optical system, using TSIM, a total of 24 and 20 discrete frequency folding centers are needed for the rectangle, Figure 2c, and rhombus, Figure 2d, tiles, respectively, with each sampled once and an additional frame containing the unperturbed baseband. The reduced number of required frames per reconstructed SR image is a direct result of the cascading reconstruction algorithm and the efficient overlap design between the folded frequencies. Defining reconstruction efficiency as the reconstructed Fourier domain unit area per the number of required frames, we obtain efficiencies of, 1.74 and 1.87 for rhombus TSIM, and rectangle TSIM, respectively, compared to 0.775 for classical MM-SIM.

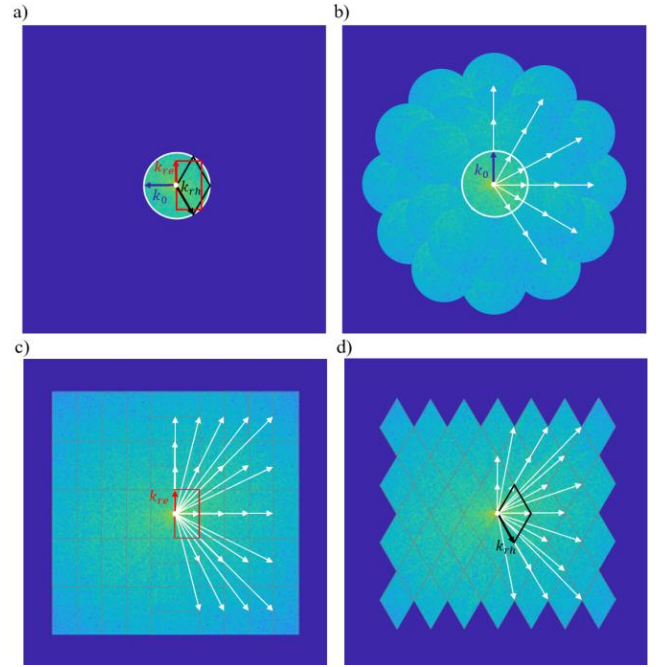

Figure 2: Fourier-domain representation of different tessellation configurations. (a) Diffraction limited system's baseband. The cutoff frequency $k_0$, bounded within a circular shape OTF. A rectangle or rhombus tile can be cropped out from within the OTF shape, with $k_{re}$ and $k_{rh}$ being the unit vectors for rectangle and rhombus tiles respectively. (b) Expansion of the spectral frequency content with linear MM-SIM. (c-d) Expansion of the spectral frequency content with rectangle and rhombus tiles, respectively. White arrows correspond to central frequencies vectors, around which spectral content is folded into the baseband of the optical system.

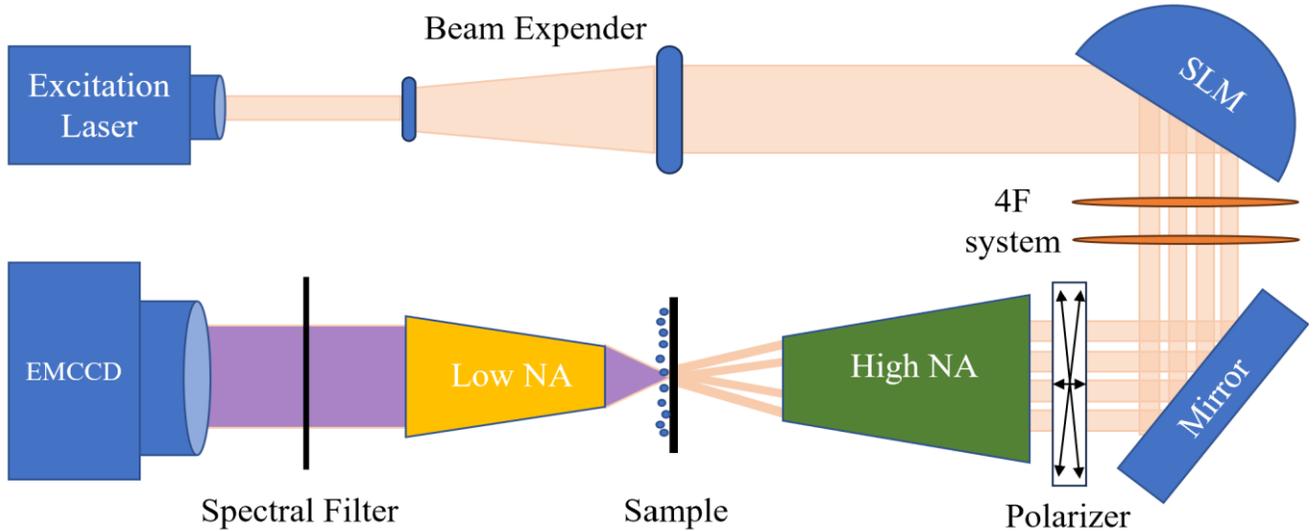

Figure 3: Schematics of TSIM experimental setup. Excitation laser is collimated and projected onto an SLM, capable of providing both phase and amplitude control over the reflected laser beam. A set of beams, spatially separated, and independently controlled, are then passed through a 4F lens system and a polarizer into the entrance pupil of high-NA objective. The beams interfere at the focal plane of the objective, coinciding with a fluorescence beads sample, creating "Moiré pattern" profile. The illumination profile excites the fluorescence beads emission response, which is then collected with a low-NA objective and imaged with an EMCCD after going through a spectral filter to separate the emission and excitation signals. For this demonstration we used a red HeNe 633 [nm] excitation laser by Newport, "Pluto" 1920X1080 SLM by Holoeye, linear polarizer, and a Nikon Plan 100X/0.90 objective for the illumination arm. For the detection arm we used upright Zeiss-axio scope as a fixture with Mitutoyo (M PLAN APO S) 50X/0.42 low-NA objective, dichroic long pass filter FEL700, and iXon Ultra 888 EMCCD, with 13μ pixel pitch, by Andor.

Resolving Fourier domain tessellation requires folding specific frequency content into the baseband of the optical system. This Frequency-folding relies on the moiré effect, which occurs when an artificial illumination profile is overlaid on the sample. The emitted signal then propagates in free space and is captured by an optical system. In Figure 2, the white arrows indicate the frequency centers around which the tiles are folded into the baseband of the optical system. Thus, these specific frequencies must be incorporated within the illumination profile. Using an SLM, we can precisely design an illumination profile to include specific frequencies, as demonstrated in our experimental TSIM setup, shown in Figure 3.

Although the SLM in our setup provides full control over the reflected planewave[22], modulating the illumination profile phase is unnecessary. This relief allows us using the first diffraction order of the SLM instead of the fourth[23], thereby increasing the reflected illumination power. By projecting an amplitude mask onto the SLM, only a spatially confined portion of the collimated laser beam is reflected. By adjusting the mask two distinct and independent collimated beams are reflected. These beams are directed onto the entrance pupil of a high-NA objective, forming a standing-wave interference pattern with a specific spectral frequency at the focal point. By adjusting the relative alignment and distance between the beams, we create illumination profiles with discrete spectral content, tiling the Fourier domain with rectangular tile shapes.

## RESULTS AND DISCUSSION

We use a total of seven illumination profiles: one with uniform planewave, two with a horizontal sinusoidal profile, two with a vertical sinusoidal profile, and two with diagonal sinusoidal profiles. By implementing two orders of cascading reconstruction, a cutoff frequency enhancement of 3-fold in the horizontal, vertical, and diagonal directions is demonstrated.

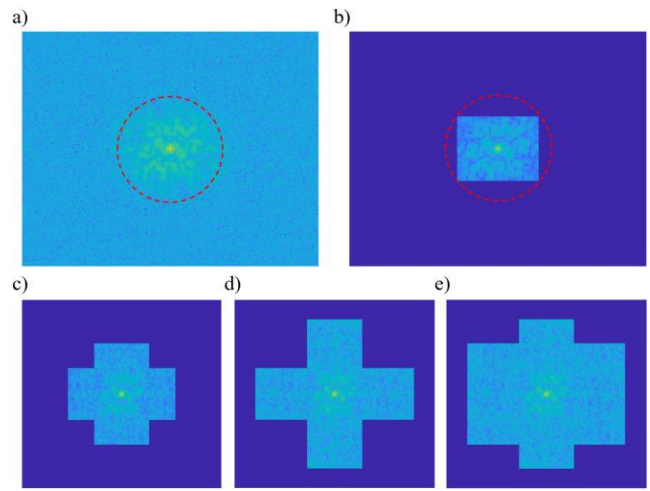

Figure 4: TSIM experimental results in Fourier domain. Diffraction limited baseband is marked with red dashed circle. (a) Diffraction limited image. (b) Cropped rectangle region used for the reconstruction scheme. (c-d) TSIM reconstruction for 1st and 2nd vertical and horizontal orders respectfully. (e) Combined horizontal, vertical and diagonal components.

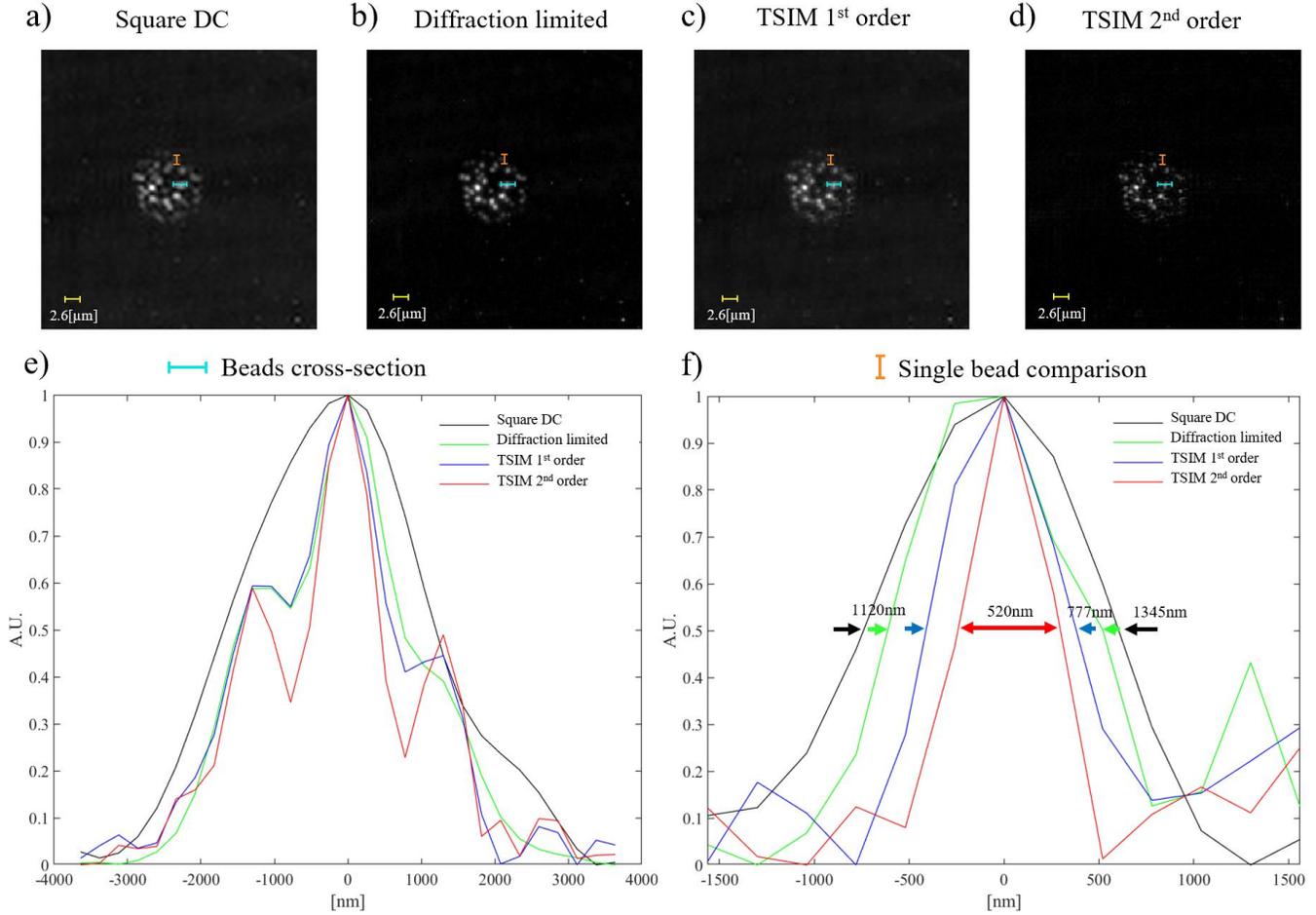

Figure 5: TSIM Experimental results. Super-resolution demonstration of fixed fluorescence beads (Spherotech, Sky Blue, FP-0570-2) with TSIM experimental setup. (a) Real domain representation of the square DC component, cropped out digitally from diffraction limited image, and used as the basis for TSIM reconstruction. (b) Diffraction limited image taken with low-NA (0.42) objective. (c) 1st order TSIM reconstruction results, based on a total of three frames. (d) 2nd order TSIM reconstruction results, based on a total of 7 frames, including diagonal components. (e) Spatial resolution increase is shown using three beads normalized cross section comparison, evolving from completely indistinguishable, through detectable in 1st order TSIM, up to fully resolved in 2nd order TSIM. (f) Normalized single bead FWHM cross-section comparison, from 1.34 [um] in the cropped rectangle DC component, through 1.1 [um] in the diffraction limited image, where 1st order TSIM reconstruction reaches 0.77 [um] and finally 0.52 [um] achieved with 2nd order TSIM reconstruction.

A comparison between the diffraction-limited image and the TSIM reconstructed image with 1st and 2nd reconstruction orders, corresponding to the first and second Fourier expansion processes, is shown for comparison in Fourier and real domains, in Figure 4 and Figure 5 respectfully.

By implementing a digital mask containing two circular bright spots, each with a 40-pixel radius, over the SLM, we selectively reflect two collimated beams onto the entrance pupil of a 0.9 NA objective. The beams form a standing-wave interference profile at the objective focal point that coincides with the fluorescence bead sample. By digitally controlling the distance and relative orientation between the two circular bright regions on the SLM, we generate a total of seven illumination profiles, one planewave and six with frequencies based on the selected folding process for rectangular tessellation. A single fluorescence transmittance frame is then captured with a 0.42 NA objective and iXon Ultra EMCCD for each illumination profile, accounting for a total of seven frames. Following the de-aliasing procedure described above, a total of six aliased spectral components are resolved and reallocated to their original position in the Fourier domain. Finally, the TSIM super-resolved image is obtained by applying the inverse Fourier transform to the reconstructed Fourier domain.

The reconstruction results shown in Figure 5 clearly demonstrate that the TSIM process successfully enhances both temporal and spatial resolution, surpassing the diffraction limit of our optical system and the temporal resolution limit of linear MM-SIM. Considering the finite size of the fluorescence beads used in this demonstration and recalling Abbe's resolution limit, we use the measured single-bead FWHM to define the optical system resolution power. Compared to the square DC image, used as the diffraction limited input for the TSIM reconstruction process, the 1st order

TSIM improves the spatial resolution by approximately 1.9-fold, while the 2nd order TSIM achieves approximately 2.6-fold improvement, as shown in Figure 5f. Compared to the non-cropped diffraction limited image we measure 1.4-fold and 2.1-fold resolution improvement for the 1st and 2nd TSIM reconstruction orders, respectfully, fully aligning with theory. Considering the Rayleigh criterion[24], alternatively, three indistinguishable beads in the square DC image, become noticeable with the 1st TSIM reconstruction order and are fully resolved with the 2nd TSIM reconstruction order, as shown in Figure 5e. The non-cropped diffraction-limited image, shown in Figure 5b, serves as the ground truth for the 1st TSIM reconstruction order, as it contains frequencies recovered by this reconstruction step, thus demonstrating the reconstruction credibility. The temporal resolution improvement is based on a total of seven frames captured for TSIM, compared to the 18 frames required for an equivalent MM-SIM reconstruction. Furthermore, the TSIM was performed without controlling or measuring the illumination profile phase.

The 2D fluorescence image shown in Figure 5Figure 5 is accompanied by its normalized 3D representation, shown in Figure 6, to better demonstrate the modulation depth difference between adjacent peaks in the diffraction-limited and TSIM reconstructed images. The normalized 3D representation provides a clear distinction in terms of the peak-to-valley modulation depth, eliminating any suspicion of apparent resolution improvement arising from the gray-level dynamic range representation. This visual comparison verifies that the super-resolved image generated with TSIM represents the physical reality truthfully.

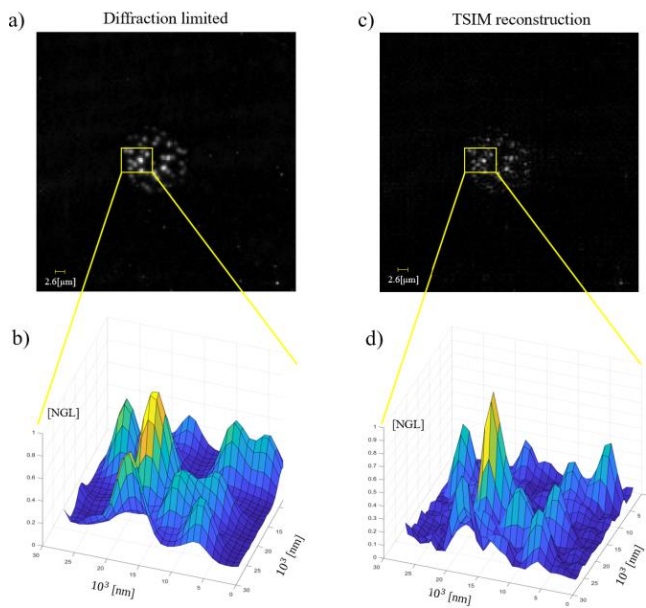

Figure 6: TSIM reconstruction VS diffraction limited image. (a) 2D diffraction limited image. (b) 3D representation of cropped region, defined by yellow rectangle. (c-d) Equivalent 2D and 3D TSIM reconstruction results. Normalized Gray Level (NGL) used for intensity indication in (b) and (d).

## CONCLUSION

We propose addressing SIM reconstruction as a Fourier-domain tessellation challenge with TSIM framework. Instead of employing the brute-force mechanism for frequency folding and unfolding, used in classical SIM, TSIM offers a fine-tuned process allowing efficient super-resolution image reconstruction with a simplified optical apparatus. Rectangle TSIM has been shown to provide approximately 2.6-fold spatial resolution enhancement, similar to MM-SIM, with a temporal resolution increase of almost 3-fold.

Both the Fourier domain tessellation paradigm and the cascading reconstruction procedure presented herein can be further incorporated into numerous previously proposed linear and nonlinear SIM architectures. Additional baseline tile shapes, and tessellation realization strategies, can be analyzed and compared to provide the long-sought for practical improvement in SRM temporal resolution. This is an important step towards achieving robust, real-time, live cell imaging beyond the diffraction limit.

## AUTHOR INFORMATION

Corresponding Author

*E-mail: guy@ee.technion.ac.il

Notes

The authors declare no competing financial interests.

## ACKNOWLEDGMENT

The authors acknowledge the fruitful discussions with S.Dolev and O.Eyal.